\documentclass[final]{JHEP3} 

\JHEPspecialurl{http://jhep.sissa.it/JOURNAL/JHEP3.tar.gz}

\usepackage{epsfig,multicol,bbm,rotating}

\newcommand\fverb{\setbox\fverbbox=\hbox\bgroup\verb}
\newcommand\fverbdo{\egroup\medskip\noindent%
			\fbox{\unhbox\fverbbox}\ }
\newcommand\fverbit{\egroup\item[\fbox{\unhbox\fverbbox}]}
\newbox\fverbbox

\def\chioi{\tilde{\chi}^0_1}

\def\sls{\tilde{l}}
\def\sql{\tilde{q}_L}
\def\sqr{\tilde{q}_R}

\def\gl{\tilde{g}}

\def\poi{\mathcal{M}_i}
\def\poo{\mathcal{M}_0}
\def\poj{\mathcal{M}_1}
\def\pojj{\mathcal{M}_2}

\def\max{{\rm max}}
\def\min{{\rm min}}
\def\MT2{M_{T2}}

\def\m0{${0}$}

\def\tq{{\tilde q}}

\title{On measuring the masses of pair-produced semi-invisibly decaying particles at hadron colliders}

\author{Daniel R. Tovey\\ Department of Physics and
	Astronomy,\\ University of Sheffield, \\ Hounsfield Road,
	Sheffield S3 7RH, UK\\ E-mail:\email{ daniel.tovey@cern.ch}}

\received{} 		
\accepted{}		

\preprint{}

\abstract{A straightforward new technique is introduced which enables
measurement at hadron colliders of an analytical combination of the
masses of pair-produced semi-invisibly decaying particles and their
invisible decay products. The new technique makes use of the
invariance under contra-linear Lorentz boosts of a simple combination
of the transverse momentum components of the aggregate visible
products of each decay chain. In the general case where the invariant
masses of the visible decay products are non-zero it is shown that in
principle the masses of both the initial particles from the hard
scattering and the invisible particles produced in the decay chains
can be determined independently. This application is likely to be
difficult to realise in practice however due to the contamination of
the final state with ISR jets. The technique may be of most use for
measurements of SUSY particle masses at the LHC, however the technique
should be applicable to any class of hadron collider events in which
heavy particles of unknown mass are pair-produced and decay to
semi-invisible final states.}


\begin{document} 


\section{Introduction}\label{sec1}
In R-Parity conserving SUSY events at hadron colliders SUSY particles
(`sparticles') must be pair-produced and undergo cascade decay to the
Lightest Supersymmetric Particle (LSP), which is often invisible and
hence a dark matter candidate. The presence of two such invisible
particles in the final state, together with imperfect detector
hermeticity close to the beam-pipe and an uncertain parton
centre-of-mass energy, prevents the use of conventional invariant mass
or transverse mass techniques for sparticle mass measurement. Similar
challenges are faced when attempting to measure the mass of any
pair-produced particles with visible and invisible decay products.

Several approaches to this general problem have been documented,
usually in the context of measuring SUSY particle masses. Given a
sufficiently long decay chain constraints on analytical combinations
of sparticle masses can be obtained from the positions of end-points
in distributions of invariant masses of combinations of visible SUSY
decay products (jets, leptons etc.)  \cite{Hinchliffe:1996iu}. Given a
number of such constraints, the system of equations may be
solved with a numerical fit to obtain the individual masses
\cite{Hinchliffe:1996iu,Allanach:2000kt,Miller:2005zp}. It was
recently shown that the mass precision obtained from this technique
can be improved by subsequently performing combined fits to individual
events, imposing both experiment end-point constraints and event
$E_T^{miss}$ constraints \cite{Nojiri:2007pq}.

When the number of kinematic end-point constraints provided by a given
decay chain is insufficient to fully constrain individual sparticle
masses alternative techniques must be employed. One possible approach
involves solving simultaneously the mass-shell conditions obtained
from several events containing the same decay chain
\cite{Kawagoe:2004rz}. This {\it mass-relation method} exploits the
small widths of SUSY states, allowing the mass of each state appearing
in the considered events to be assumed to be constant. 

A second approach to this problem is to select events in which the
same decay chain appears in both `legs' of each selected event. In
this case additional constraints are provided by the components of the
event ${\bf E_T}^{miss}$ vector\footnote{We denote three-vector and
two-vector quantities with {\bf bold} case, while the corresponding
magnitudes are denoted with standard case. Four-vector quantities are
written in standard case.}, and again use can be made of the sparticle
narrow-width approximation to equate the masses in the two legs. This
permits the construction of further distributions with kinematic
end-points related to the masses of sparticles present in the event.

One example of a technique of this kind is the {\it stransverse mass}
method \cite{Lester:1999tx,Allanach:2000kt,Barr:2003rg}. Consider two
identical heavy SUSY states $\delta_1$ and $\delta_2$, decaying
respectively to visible products $v_1$ and $v_2$ and identical lighter
states $\alpha_1$ and $\alpha_2$. If ${\bf p_T}(\alpha_1)$, the
transverse momentum vector of $\alpha_1$, were known then it would be
possible to calculate $m_T(\delta_1)$, the transverse mass of
$\delta_1$, which is bounded from above by $m(\delta)$. If ${\bf
p_T}(\alpha_1)$ is known however then so is ${\bf p_T}(\alpha_2)$
through application of the event ${\bf E_T}^{miss}$
constraints. Therefore $m_T(\delta_2)$ could also be calculated, a
quantity which must also be less than $m(\delta)$. Consequently the
maximum value of $m_T(\delta_1)$ and $m_T(\delta_2)$ provides a
variable with an end-point whose position measures $m(\delta)$. Of
course in reality we are not able to measure ${\bf p_T}(\alpha_1)$ or
${\bf p_T}(\alpha_2)$ however the great insight of
Ref.~\cite{Lester:1999tx} was the realisation that if we can find a
test value ${\bf p_T}(\alpha_1)$ which minimises this maximum
transverse mass, we can be sure that the minimised-maximised
transverse mass is also bounded from above by $m(\delta)$. This
`minimax' transverse mass quantity is referred to as the `stransverse
mass' or $M_{T2}$.

The development of the stransverse mass technique was particularly
important because for the first time it allowed the measurement of
masses of sparticles decaying through very short cascades, for
instance $\sqr \rightarrow q \chioi$ or $\sls \rightarrow l
\chioi$. Furthermore an analytical expression for $M_{T2}$ has
recently been derived, valid in cases where the centre-of-mass (CoM)
frame is at rest in the laboratory transverse plane
\cite{Lester:2007fq}, thus simplifying its use considerably. The
technique inherits one draw-back from its use of the transverse masses
of $\delta$ decay products however, namely that it requires the use of
$m(\alpha)$ as an input. $M_{T2}$ may therefore be described more
correctly as an ensemble of variables, one for each assumed value for
the unknown quantity $m(\alpha)$. The dependence of $M_{T2}$ on
$m(\alpha)$ has been determined to be approximately
$m(\delta)-m(\alpha)$ in specific cases \cite{Barr:2003rg}, however it
would in general be preferable if the definition of the variable were
independent of the unknown quantities to be measured. In that case the
mass constraints obtained from an end-point fit would be uncorrelated
with other measurements and hence could be used as input to a global
mass fit.

In this paper we will propose a very simple technique which seeks to
address the same problem as the stransverse mass technique, but which
approaches the problem from a different perspective. The new technique
will allow a simple analytical combination of particle/sparticle
masses to be constrained in a precise and model-independent
manner. Furthermore the technique will offer at least in principle the
prospect of measuring individual particle masses, as postulated for
the stransverse mass technique in
Refs.~\cite{Cho:2007qv,Gripaios:2007is,Barr:2007hy,Cho:2007dh}. The
new technique will be applicable to any class of events in which heavy
particles of unknown mass are pair-produced and decay to
semi-invisible final states.

The structure of the paper is as follows. Section~\ref{sec2} will
describe the principles underlying the technique and investigate the
properties of the new variable upon which it is based.
Section~\ref{sec4} will illustrate application of the technique to the
problem of constraining sparticle mass combinations with $\tilde{q}_R
\rightarrow q \chioi$ pair events at the LHC. Section~\ref{sec5} will
outline extension of the technique to measurement of individual
particle masses. Section~\ref{sec6} will conclude and discuss avenues
for future work.

\section{Description of technique}
\label{sec2}
\subsection{Background}
\label{subsec2.1}

Consider `symmetric' events in which identical cascade decay
chains of the form
\begin{equation}
\label{eqn2}
\delta \rightarrow \alpha  v
\end{equation}
occur in each leg $i$ of the event. We shall refer to the initial
particles produced in the hard scattering as $\delta_i$. We shall
further consider $n$ step decay chains in each leg consisting of $n-1$
decays, such that the $(n-1)^{th}$ decays produce invisible particles
$\alpha_i$. The visible products of decays $1$ to $(n-1)$ in each leg
will be considered as single systems $v_i$ of mass $m(v_i)$ and
four-momentum $p(v_i)$. We shall assume that no invisible particles
other than $\alpha$ are produced in the decay chains. The particles
$\delta_i$ and $\alpha_i$ have common masses which are respectively
$m(\delta)$ and $m(\alpha)$.

This parameterisation of the decay chains is quite general. The case
$n=2$ corresponds to SUSY chains such as $\sqr \rightarrow q \chioi$
or $\sls \rightarrow l \chioi$, with $\alpha$ identified as the LSP
$\chioi$. In these cases $m^2(v_i) << p^2(v_i)$ at the LHC. For longer
SUSY chains we can choose the number of decays provided we can
unambiguously identify the visible products of those decays. If $n$ is
equal to the total number of sparticles in the chain then $\alpha$ is
again the LSP. For chains with $n>2$ steps the distributions of
invariant masses $m(v_i)$ can display kinematic end-points sensitive
to analytical combinations of sparticle masses appearing in the chain
\cite{Hinchliffe:1996iu}. This information is used to constrain the
individual masses in the end-point method but will be incidental to
the technique described here.

Consider now the use of $N$ symmetric events of the general form of
Eqn.~\ref{eqn2} to measure $m(\delta)$ and $m(\alpha)$. This
problem reduces to one of solving 6$N$ non-linear simultaneous equations,
with each event providing the mass-shell conditions:
\begin{eqnarray}
[p(v_1)+p(\alpha_1)]^2 & = & [p(v_2)+p(\alpha_2)]^2 = m^2(\delta),\cr
[p(\alpha_1)]^2 & = & [p(\alpha_2)]^2 = m^2(\alpha),
\label{eqn3}
\end{eqnarray}
together with two ${\bf E_T}^{miss}$ constraints:
\begin{eqnarray}
        p_x(\alpha_1)+p_x(\alpha_2) & = & E_x^{miss}, \cr
        p_y(\alpha_1)+p_y(\alpha_2) & = & E_y^{miss}.
\label{eqn4}
\end{eqnarray}
Each event contributes 2 unknown masses, which are common to all
events, and 8 unknown $\alpha_i$ four-momentum components, which
differ between events. The total number of unknown parameters is
therefore $8N+2$ while the number of constraints is $6N$ and so the
system of equations is highly under-constrained.

It may seem surprising at first that the above system of equations can
be solved at all, however it should be noticed that we are not
concerned with measuring the four-momenta $p(\alpha_1)$ and
$p(\alpha_2)$ for all events, but rather with measuring only the
common masses $m(\delta)$ and $m(\alpha)$ using at least one
event. Consequently we may set out to discard events in which the
unknown masses depend on unknown four-momentum components. The problem
therefore reduces to one of finding variables dependent only on the
measurable quantities $p(v_1)$, $p(v_2)$ and ${\bf E_T}^{miss}$ which
identify events where the masses also depend only on those measurable
quantities. This general approach is effectively that taken by
kinematic end-point techniques, in which the variables identifying the
events, such as $m(ll)$ or $M_{T2}$, are also those which provide the
mass measurment. This is also the approach which shall be taken here.

\subsection{Transverse momentum end-points}
\label{subsec2.2}

One possible starting point for this problem was outlined in
Ref.~\cite{Tovey:2000wk}. In an effective two-body decay process of
the type considered above the magnitude of the three-momentum of the
visible decay products in the rest frame of $\delta_i$ is given by
\begin{eqnarray}
\label{eqn5}
|{\bf p}(v_i)| & = & \frac{1}{2} \frac{\sqrt{[m^2(\delta) - m^2(\alpha) + m^2(v_i)]^2 - [2m(\delta)m(v_i)]^2}}{m(\delta)} \cr 
& \equiv & \frac{1}{2} \poi,
\end{eqnarray}
which defines the 2-body mass parameter $\poi$. It will also be
useful for the discussion which follows to define the equivalent
quantity $\poo$ for the special case where $m(v_i)=0$:
\begin{equation}
\label{eqn5a}
\poo \equiv \frac{m^2(\delta) - m^2(\alpha)}{m(\delta)}.
\end{equation}
If $\delta_i$ has a small boost in the laboratory transverse frame,
then the laboratory transverse momentum of $v_i$ is of order
$\poi/2$. This dependence of the momenta of visible decay products on
the masses of heavy particles further up the decay chain is the reason
that variables such as the `effective mass' \cite{Hinchliffe:1996iu}
used in SUSY studies are sensitive to such masses.

In principle we can improve on the use of \emph{ad hoc} variables such
as the effective mass however. In the rest frame of $\delta_i$ the
magnitude of the momentum of $v_i$ transverse to the beam direction
\footnote{We denote quantities measured in the $\delta_1\delta_2$
CoM frame with primed variables and those measured in
the rest frames of $\delta_1$ or $\delta_2$ with unprimed variables.},
$p_T(v_i)$, is bounded from above by $\poi/2$ because
\begin{equation}
p_T(v_i) = \frac{\poi}{2} \sin \psi_i,
\label{eqn6}
\end{equation}
where $\psi_i$ is the polar decay angle relative to the beam
direction. Consequently if we could measure $p_T(v_i)$ we could
constrain the masses. 

Unfortunately however we are not able to measure $p_T(v_i)$
directly -- instead we measure the equivalent quantity in the
laboratory frame: $p_T'(v_i)$. To proceed further we assume that
the $\delta_1 \delta_2$ CoM frame is at rest in the
laboratory transverse plane. This condition can be enforced by
selecting events in which the net transverse momentum of the final
state excluding the $\delta_1$ and $\delta_2$ decay products is
small. In this case $p_T'(v_i)$ is related to $p_T(v_i)$ by a proper
Lorentz transformation in the transverse plane through the well-known
relation:
\begin{equation}
p_T'^2(v_i) = \frac{1}{1-\beta^2}\left [p_T(v_i) \cos\phi_i
\pm \beta E(v_i) \right ]^2 + p_T^2(v_i)\sin^2\phi_i,
\label{eqn7}
\end{equation}
where $\beta$ is the transverse boost factor ($0<\beta<1$), $E(v_i)$
is the energy of $v_i$ and $\phi_i$ is the angle in the rest frame of
$\delta_i$ between the boost direction and ${\bf p_T}(v_i)$. For given
$v_i$ we know neither $\beta$ nor $\phi_i$ and hence we are not able
to reconstruct $p_T(v_i)$. Nevertheless we do know from conservation
of momentum that in the $\delta_1 \delta_2$ CoM frame, and hence the
laboratory transverse plane, the boost applied to $v_2$ is equal and
opposite to that applied to $v_1$.

To proceed further we shall attempt to find a quantity which can be
calculated from the components of ${\bf p_T}(v_1)$ and ${\bf
p_T}(v_2)$ which remains unchanged if calculated with the
corresponding components of ${\bf p_T'}(v_1)$ and ${\bf
p_T'}(v_2)$. If we could find such a quantity then we could use it
to relate momenta measured in the $\delta_1 \delta_2$ CoM frame to
those measured in the $\delta_1$ and $\delta_2$ rest frames and hence
constrain $\poi$.

\subsection{Cotransverse mass and contransverse mass}
\label{subsec2.3}

Consider first a system containing two particles $v_1$ and $v_2$ with
masses $m(v_1)$ and $m(v_2)$ measured in some frame F$_{(0)}$ to have
four-momenta $p(v_1)$ and $p(v_2)$. If both these particles are now
measured in a different frame F$_{(1)}$ it is well known that the mass
obtained from $p(v_1)+p(v_2)$ remains unchanged, i.e. the quantity
\begin{eqnarray}
\label{eqn8}
m^2(v_1,v_2) & = &[E(v_1)+E(v_2)]^2 - [{\bf p}(v_1)+{\bf p}(v_2)]^2 \cr
& = & m^2(v_1) + m^2(v_2) + 2[E(v_1)E(v_2)-{\bf p}(v_1) \cdot {\bf p}(v_2)]
\end{eqnarray}
is invariant. Another way to interpret this is that when particles
$v_1$ and $v_2$ are subjected to co-linear boosts of equal magnitude
$m^2(v_1,v_2)$ is invariant.

Now let us examine what happens when we start from one frame
F$_{(0)}$, but boost particles $v_1$ and $v_2$ to different frames
F$_{(1)}$ and F$_{(2)}$ respectively. These new frames are
distinguished by the fact that their boosts are of equal magnitude but
{\it opposite} direction in frame F$_{(0)}$. In other words particles
$v_1$ and $v_2$ are subjected to contra-linear boosts of equal
magnitude. Clearly $m^2(v_1,v_2)$ is no longer an invariant -- this
can be seen for instance by considering ${\bf p}(v_1)=-{\bf p}(v_2)$
in which case $E(v_1)+E(v_2)$ increases with increasing $\beta$ while
${\bf p}(v_1)+{\bf p}(v_2)$ remains zero.

Consider now a new quantity $M_C$ equivalent to the invariant mass
obtained from $p(v_1)+\mathcal{P}(p(v_2))$ where $\mathcal{P}$ is the
standard parity transformation operator:
\begin{eqnarray}
\label{eqn9}
M_C^2(v_1,v_2) & \equiv & [E(v_1)+E(v_2)]^2 - [{\bf p}(v_1)-{\bf
p}(v_2)]^2 \cr & = & m^2(v_1) + m^2(v_2) + 2[E(v_1)E(v_2)+{\bf p}(v_1)
\cdot {\bf p}(v_2)].
\end{eqnarray}
This quantity is invariant under the contra-linear boosts
considered above. Denoting quantities measured in F$_{(0)}$ with
primed variables, and those measured in F$_{(1)}$ and F$_{(2)}$ with
unprimed variables, and defining the $\hat{x}$ direction to be the
boost direction, this can easily be demonstrated:
\begin{eqnarray}
\label{eqn10}
M_C'^2(v_1,v_2) & = &  [E'(v_1)+E'(v_2)]^2 - [{\bf p}'(v_1)-{\bf p}'(v_2)]^2 \cr\cr
& = & \mbox{ }\mbox{ }\gamma^2 \left [ E(v_1)+\beta
p_x(v_1)+E(v_2)-\beta p_x(v_2) \right ]^2 \cr &  & - \gamma^2 \left [ p_x(v_1)+\beta
E(v_1)-p_x(v_2)+\beta E(v_2) \right ]^2 \cr &  & - \left [p_y(v_1)-p_y(v_2)
\right] ^2 - \left [p_z(v_1)-p_z(v_2) \right] ^2 \cr\cr
& = & \mbox{ }\mbox{ }\gamma^2 \left ( [E(v_1)+E(v_2)]^2 + \beta^2
[p_x(v_1)-p_x(v_2)]^2 + 2\beta[E(v_1)+E(v_2)][p_x(v_1)-p_x(v_2)] \right) \cr &  & - \gamma^2 \left (
\beta^2[E(v_1)+E(v_2)]^2  + [p_x(v_1)-p_x(v_2)]^2 +
2\beta[E(v_1)+E(v_2)][p_x(v_1)-p_x(v_2)] \right ) \cr &  & - \left [p_y(v_1)-p_y(v_2)
\right] ^2 - \left [p_z(v_1)-p_z(v_2) \right] ^2 \cr\cr
& = & \gamma^2 \left ( [E(v_1)+E(v_2)]^2 [1-\beta^2] -
[p_x(v_1)-p_x(v_2)]^2 [1-\beta^2] \right ) \cr &  & - \left [p_y(v_1)-p_y(v_2) \right]
^2 - \left [p_z(v_1)-p_z(v_2) \right] ^2 \cr\cr
& = & [E(v_1)+E(v_2)]^2 - [{\bf p}(v_1)-{\bf p}(v_2)]^2 \cr\cr
& = & M_C^2(v_1,v_2).
\end{eqnarray}
Since $M_C(v_1,v_2)$ is invariant under contra-linear boosts of equal
magnitude its value can be calculated from the momenta and energies of
$v_1$ and $v_2$ in any pair of frames F$_{(1)}$ and F$_{(2)}$ related
to F$_{(0)}$ by such boosts. For instance in the case considered above
F$_{(0)}$ could be identified with the $\delta_1 \delta_2$ CoM frame
and F$_{(1)}$ and F$_{(2)}$ identified with the rest frames of
$\delta_1$ and $\delta_2$, in which $|{\bf p}(v_1)|=\poj/2$ and $|{\bf
p}(v_2)|=\pojj/2$.

From a practical perspective the quantity $M_C(v_1,v_2)$ defined by
Eqn.~\ref{eqn9} is relevant only to cases where the $\delta_1
\delta_2$ CoM frame is at rest in the laboratory frame, for instance
in collisions at a lepton collider such as LEP or the
ILC. At a hadron collider the scenario is more complicated. As
discussed above, co-linear boosts in the laboratory transverse plane
can be limited by selecting events in which the net transverse
momentum of the final state excluding the $\delta_1$ and $\delta_2$
decay products is small. There remains however a potentially large
co-linear boost in the beam ($\hat{z}$) direction caused by the
differing proton momentum fractions of the colliding partons in the
event initial state. $M_C(v_1,v_2)$ is not invariant under co-linear
boosts of $v_1$ and $v_2$ because $\mathcal{P}$ does not commute with
proper Lorentz transformations. Consequently we must focus purely on
quantities constructed from momentum components measured in the
laboratory plane transverse to the beam direction.

If $v_1$ and $v_2$ were subjected to co-linear rather than
contra-linear equal magnitude boosts in the laboratory transverse
plane then a suitable invariant quantity to consider would be the
transverse mass $m_T(v_1,v_2)$ \cite{Arnison:1983rp}, hereafter
refered to as the {\it cotransverse mass}. $m_T(v_1,v_2)$ is defined
by:
\begin{eqnarray}
\label{eqn11}
m_T^2(v_1,v_2) & = &[E_T(v_1)+E_T(v_2)]^2 - [{\bf p_T}(v_1)+{\bf p_T}(v_2)]^2 \cr
& = & m^2(v_1) + m^2(v_2) + 2[E_T(v_1)E_T(v_2)-{\bf p_T}(v_1) \cdot {\bf p_T}(v_2)],
\end{eqnarray}
where 
\begin{equation}
\label{eqn12}
E_T(v_i) = \sqrt{p_T^2(v_i)+m^2(v_i)}.
\end{equation}
This quantity is useful because it is bounded from above by
$m(v_1,v_2)$. When $m(v_1)=m(v_2)=0$ the following simplification can
be made:
\begin{equation}
\label{eqn13}
m_T^2(v_1,v_2) = 2p_T(v_1)p_T(v_2)(1-\cos \phi_{12}),
\end{equation}
where $\phi_{12}$ is the angle between $v_1$ and $v_2$ in the
transverse plane. This illustrates that events saturating the bound on
the (co)transverse mass typically require that $v_1$ and $v_2$ be
back-to-back.

In the case of contra-linear equal magnitude boosts considered above
the equivalent quantity to the (co)transverse mass can be derived from
Eqn.~\ref{eqn9}:
\begin{eqnarray}
\label{eqn14}
M_{CT}^2(v_1,v_2) & \equiv & [E_T(v_1)+E_T(v_2)]^2 - [{\bf p_T}(v_1)-{\bf p_T}(v_2)]^2\cr
& = & m^2(v_1) + m^2(v_2) + 2[E_T(v_1)E_T(v_2)+{\bf p_T}(v_1) \cdot {\bf p_T}(v_2)].
\end{eqnarray}
We shall refer to this quantity as the {\it contransverse mass}. This
has the property that when $m(v_1)=m(v_2)=0$ it reduces to
\begin{equation}
\label{eqn15}
M_{CT}^2(v_1,v_2) = 2p_T(v_1)p_T(v_2)(1+\cos \phi_{12}),
\end{equation}
where if $p_T(v_1)$ and $p_T(v_2)$ are measured in the laboratory
transverse plane then $\phi_{12}$ is the angle between $v_1$ and $v_2$
in that plane. 

It is interesting to note at this point that when $v_1$ and $v_2$ are
massless and the $\delta_1 \delta_2$ CoM frame is at rest in the
laboratory transverse plane the ${\bf E_T}^{miss}$ vector can be
represented under a change of basis involving $M_{CT}(v_1,v_2)$ :
\begin{equation}
{\bf E_T}^{miss}=\{-p_x(v_1)-p_x(v_2),-p_y(v_1)-p_y(v_2)\} \rightarrow
\{p_T(v_1)-p_T(v_2),M_{CT}(v_1,v_2)\}.
\label{eqn15a}
\end{equation}
In the new basis the first component can be interpreted as the
contribution to $E_T^{miss}$ from $p_T$ asymmetry, while the second,
containing the geometric mean of $p_T(v_1)$ and $p_T(v_2)$, can be
interpreted as the contribution from event topology. In this case
$E_T^{miss}$ is given by:
\begin{equation}
E_T^{miss} = \sqrt{[p_T(v_1)-p_T(v_2)]^2 + M_{CT}^2(v_1,v_2)}.
\label{eqn15b}
\end{equation}

The physical interpretation of the contransverse mass is more
difficult than in the (co)transverse case. $M_{CT}(v_1,v_2)$ does not
represent the mass of a particle decaying to produce $v_1$ and
$v_2$. Nevertheless we expect its distribution to display an
end-point because it can in principle be calculated from the momenta
of visible decay products measured in the rest frames of $\delta_1$
and $\delta_2$, and we know from Section~\ref{subsec2.2} that these
momenta are bounded from above by $\poi/2$. For instance if
$m(v_1)=m(v_2)=0$ then $M_{CT}(v_1,v_2)$ takes a maximum value of
$\poo$, i.e.
\begin{equation}
\label{eqn16}
M_{CT}^{max} = \frac{m^2(\delta)-m^2(\alpha)}{m(\delta)}.
\end{equation}
Interestingly this bound is saturated when $v_1$ and $v_2$ are
co-linear, in contrast to the case for the (co)transverse mass.

To summarise, we have now found a quantity bounded from above by an
analytical combination of particle masses, and which can be calculated
using momenta of visible decay products measured in the laboratory
transverse plane. We shall now consider as a use-case the practical
application of this variable to LHC data in order to measure SUSY
particle masses.

\section{Example: $\sqr\sqr$ events at the LHC}
\label{sec4}
To illustrate the application of the contransverse mass end-point
technique to LHC data, a Monte Carlo simulation study was carried out
aimed at measuring $M_{CT}^{max}$ for $\sqr$ pair production events
where each $\sqr$ decays to a quark and a $\chioi$. Squark mass
measurement in this channel using the stransverse mass method was
first studied in Ref.~\cite{Weiglein:2004hn}. The experimental
signature of this process is the presence of events with exactly two
jets and large $E_T^{miss}$. In the context of the decay chain
discussed in Section~\ref{subsec2.1} the $\sqr$ plays the role of $\delta$
and $\chioi$ that of $\alpha$. We assume that the quark jet decay
products are massless and hence Eqn.~\ref{eqn16} allows us to measure
an analytical combination of sparticle masses by measuring
$M_{CT}^{max}$.

A sample of 480k SUSY signal events equivalent to 10 fb$^{-1}$ of data
was generated from the SPS1a benchmark mSUGRA model
\cite{Weiglein:2004hn} with {\tt HERWIG 6.5}
\cite{Corcella:2000bw,Moretti:2002eu} and passed to a generic LHC
detector simulation \cite{RichterWas:2002ch} modified to impose an
80\% efficiency for electron identification, with mis-identified
electrons being added to the list of jets if $p_T(e)>10$ GeV. The {\tt
ISASUGRA 7.69} RGE code \cite{Paige:2003mg} was used to calculate the
input SUSY mass spectrum, giving $m(\sqr) \sim 548$ GeV and
$m(\chioi)=96$ GeV and hence $M_{CT}^{max}=531$ GeV. A fully inclusive
sample of SUSY events was generated in order to model SUSY
backgrounds.

Events were selected with the following requirements (with $j_i$ used
to denote jet $i$):
\begin{itemize}
\item $n_{jet}=2$ for $\Delta R = 0.4$ cone jets with $p_T(j)>10$ GeV
and $|\eta|<5.0$,
\item $n_{lep}=0$ for isolated leptons (electrons or muons) with
$p_T>5$ GeV (electrons) or $p_T>6$ GeV (electrons), $|\eta|<2.5$,
minimum $\Delta R$ with nearest jet of 0.4 and maximum energy
deposition of 10 GeV in a $\Delta R = 0.2$ isolation cone,
\item $\min[p_T(j_1),p_T(j_2)]>100$ GeV, 
\item $E_T^{miss}>200$ GeV,
\item in order to limit boosts of the $\sqr\sqr$ CoM frame in the
laboratory transverse plane, measured $p_T$ of the $j_1j_2+E_T^{miss}$
CoM frame in the laboratory transverse plane must satisfy
\begin{equation}
\label{eqn20}
\sqrt{[p_x(j_1)+p_x(j_2)+E_x^{miss}]^2+[p_y(j_1)+p_y(j_2)+E_y^{miss}]^2}<20 \mathrm{GeV},
\end{equation} 
\item $M_{CT}>200$ GeV.
\end{itemize}
The hard $p_T(j)$ and $E_T^{miss}$ cuts additionally ensure that
events easily pass typical LHC high level jet + $E_T^{miss}$ trigger
criteria such as $p_T(j)>70$ GeV and $E_T^{miss}>70$ GeV
\cite{atlas:2003hw}.

After application of these cuts many Standard Model (SM) backgrounds are
heavily suppressed:
\begin{itemize}
\item QCD jet backgrounds, while possessing a very large
cross-section, are suppressed by the $M_{CT}$ cut which rejects events
with back-to-back jets. Jet energy mis-measurement mainly
generates $E_T^{miss}$ through the first term in Eqn.~\ref{eqn15b} and
the effect on $M_{CT}$ is smaller. In order to pass the $M_{CT}$ cut
at least one high $p_T$ jet must be completely missed by the
detector. The $M_{CT}$ cut is strongly correlated with the
$D_{\pi\pi}$ variable used at the Tevatron to separate SUSY signal
from QCD backgrounds in multijet+$E_T^{miss}$ searches
\cite{Abbott:1999xc}. The fast detector simulation used in this study
is not expected to model the catastrophic loss of jets accurately, but
we do not expect this background to be dominant even when using a more
realistic simulation. In particular such events can in principle be
removed with `event cleaning' cuts, for instance by reconstructing
jets from charged particle tracks. Consequently QCD jet backgrounds
are not considered further here.
\item Hadronic or semi-leptonic $t\bar{t}$ backgrounds are suppressed
by the jet multiplicity cuts, while fully leptonic events in which
both leptons are lost inside the jets (the worst case scenario
kinematically) possess $M_{CT}$ values less than $m(t)\sim 172$ GeV,
which is the value expected for top quarks decaying to a neutrino plus
massless visible decay products. Such events therefore fail the
$M_{CT}$ cut.
\item $W+1$ jet backgrounds in which the $W$ decays to a hadronic
$\tau$ or electron faking a jet can mimic 2-jet events. Events
with large $M_{CT}$ typically possess two co-linear
jets. When the lepton is emitted co-linearly with the
initial jet its transverse momentum is given by
\begin{equation}
\label{eqn19}
p_T= \frac{1}{2} m(W) \gamma (1-\beta),
\end{equation}
where $\beta$ is the boost of the $W$ in the transverse plane. The
maximum value of $M_{CT}$ generated by this configuration is obtained
in the limit $\beta \rightarrow 1$, when $M_{CT}^{max}=m(W)$. Such
events therefore also fail the $M_{CT}$ cut and are not considered
further here.
\end{itemize}

The remaining SM backgrounds are dominated by $Z(\rightarrow\nu\nu)+2$
jets and $W(\rightarrow l\nu)+2$ jets events, where in the latter case
the lepton momentum is anti-parallel to the $W$ momentum and is
`red-shifted' such that its magnitude is below the lepton
identification $p_T$ threshold. These backgrounds were modelled with
{\tt ALPGEN} \cite{Mangano:2002ea} coupled to {\tt HERWIG 6.5}. In
order to add realism to the analysis the backgrounds were estimated
using data-driven techniques applied to the Monte Carlo `data'. The
$Z(\rightarrow\nu\nu)+2$ jets background was estimated by selecting
$Z(\rightarrow ll)+2$ jets events with similar cuts to those listed
above, but replacing the lepton veto and $E_T^{miss}$ requirements
with a requirement for two opposite-sign same-flavour leptons with
$|m(ll)-m(Z)|<10$ GeV and $|{\bf p_T}(ll)+{\bf E_T}^{miss}| > 200$
GeV. The $W(\rightarrow l\nu)+2$ jets background was estimated by
selecting $W(\rightarrow l\nu)+2$ jets events in which the lepton was
boosted and the neutrino de-boosted such that $p_T(l)> 200$ GeV and
$E_T^{miss} < 10$ GeV. Each data-driven estimate was normalised
separately to the respective Monte Carlo background $p_T(j)$
distribution below the expected SUSY signal region. In practice the
relative normalisation of the $W(\rightarrow l\nu)+2$ jets estimate
could be obtained from data, for instance with a fit to the lepton
$p_T$ spectrum in $W(\rightarrow l\nu)+2$ jets events.

SUSY backgrounds to $\sqr\sqr$ events arise primarily from processes
in which at least one $\sql$ decays through a chain producing multiple
invisible final state particles. One possible example involves
sneutrinos decaying to neutrinos and $\chioi$. In these cases the mass
of each SUSY state produced in association with the jet in the decay
of each $\sql$ is greater than that of the $\chioi$ produced in the
decay of $\sqr$ and consequently these events possess $M_{CT}^{max}$
values below those for $\sqr\sqr$. At parton level if SUSY background
events are to exceed the expected end-point, assuming correct
assignment of decay products to SUSY decay chains, then the mass of
the initially produced sparticles must be greater than $m(\sqr)$. The
main candidate for this is $\gl$ pair production in which each gluino
decays to co-linear jets in association with a $\chioi$. This process
should generate a $M_{CT}$ distribution with an endpoint at
$M_{CT}^{max}=597$ GeV for a $\gl$ mass of 612 GeV at SPS1a.

\FIGURE[ht]{
\epsfig{file=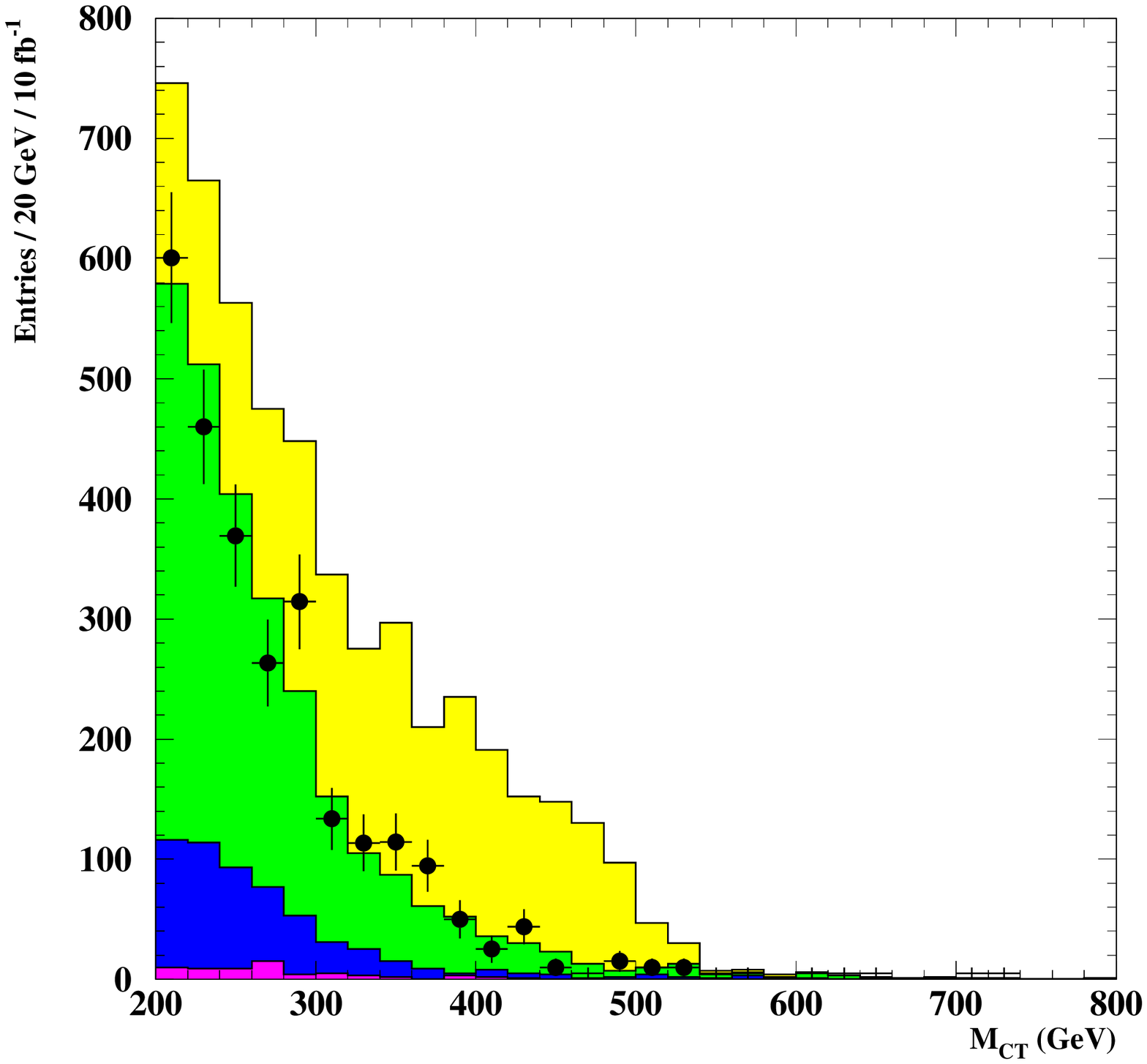,height=2.9in}
\epsfig{file=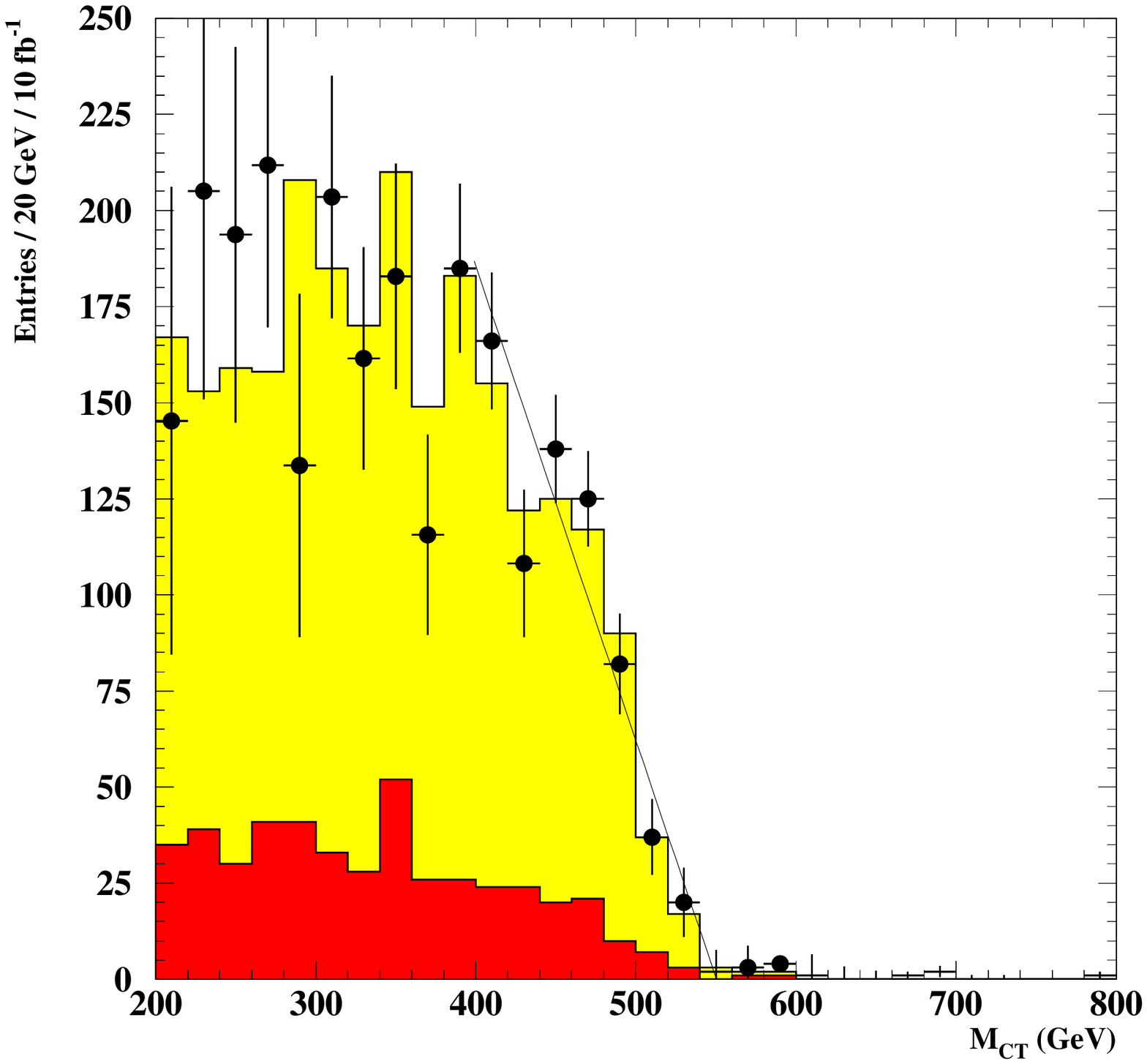,height=2.9in}
\caption{\label{fig4-4} Distributions of $M_{CT}$ values. The
left-hand figure shows the cumulative `data' distributions summing
SPS1a SUSY events (light/yellow histogram), $Z(\rightarrow\nu\nu)+2$
jets background events (medium/green), $W(\rightarrow l\nu)+2$ jets
background events (dark/blue) and diboson, top-pair and single-top
background events (magenta/medium-dark). Data-points indicate the
result of the data-driven estimate described in the text. The
data-points in the figure on the right show the result of subtracting
the data-driven estimate from the `data' distribution, with the
light/yellow histogram representing the SUSY distribution with no SM
background added. The dark/red histogram shows the contribution from
non-$\sqr \sqr$ SUSY events. The result of a simple linear end-point
fit to the data-points is shown.}}

The $M_{CT}$ distribution for events satisfying the selection cuts is
shown in Fig.~\ref{fig4-4}(left) indicating an excess of events at
large $M_{CT}$ values due to SUSY processes. As expected the
contribution from $t\bar{t}$ events (modeled with {\tt HERWIG 6.5}) is
small, as are the contributions from $WW$, $WZ$, $ZZ$ and single-top
production (also modeled with {\tt HERWIG 6.5}).  The data-points in
Fig.~\ref{fig4-4}(right) represent the same distribution after
subtracting the data-driven background estimate. As expected, a
prominent end-point feature is visible at around 530 GeV. A simple
linear fit to the endpoint determines its position to be 550 $\pm$ 53
GeV (10\% uncertainty). Use of a more sophisticated fitting function
would undoubtedly improve this precision significantly. There is also
some evidence in Fig.~\ref{fig4-4}(right) for a small excess of events
beyond the expected $\sqr\sqr$ end-point. Examination of the Monte
Carlo truth record indicates that the large $M_{CT}$ values of these
events originate either from jet mis-measurement in $\sqr\sqr$ events
or from both jets originating from the same SUSY decay chain in
non-$\sqr\sqr$ events. No $\gl\gl$ events were observed to contribute
to this region for this SPS1a SUSY model.

\FIGURE[ht]{
\epsfig{file=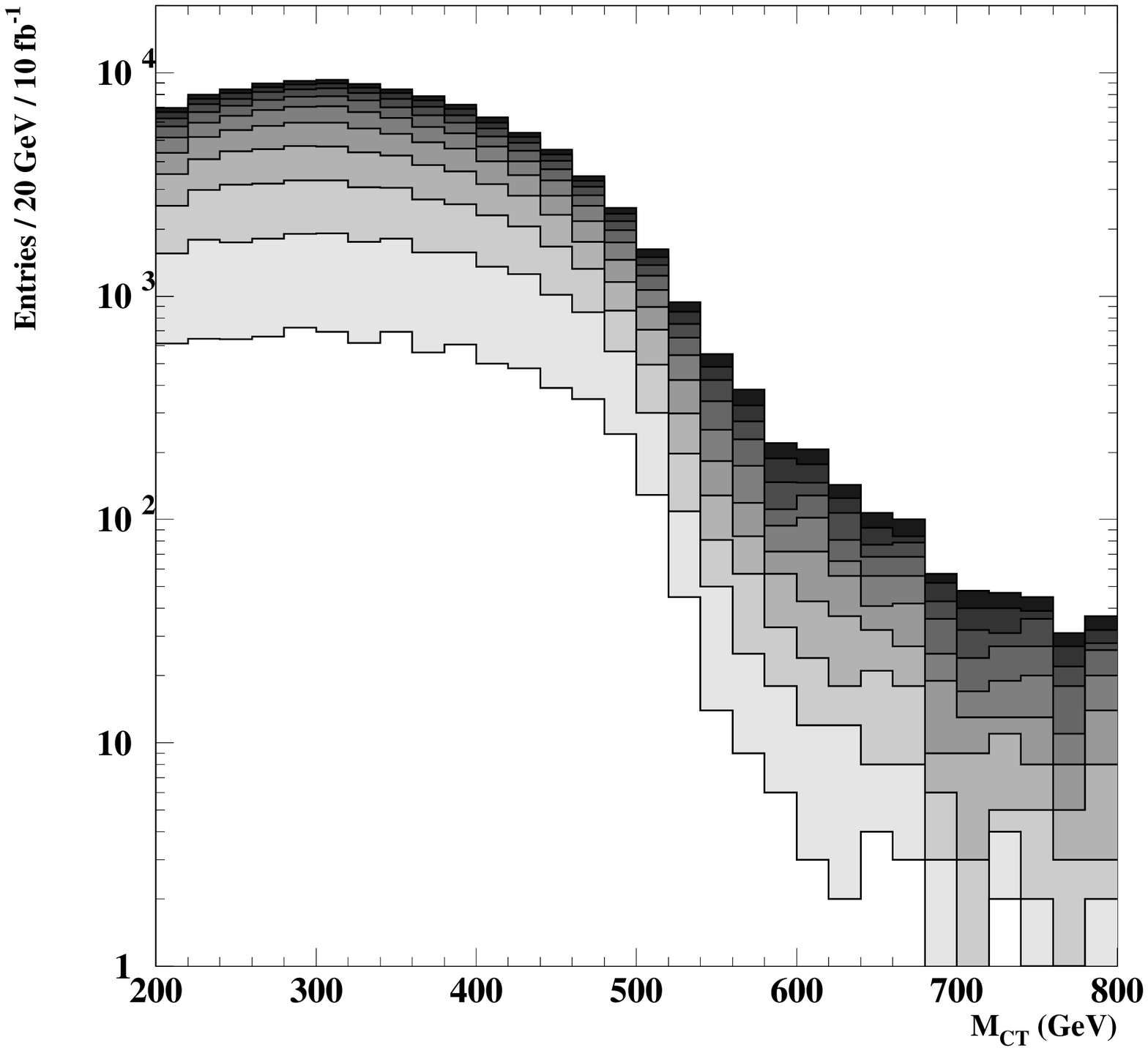,height=3.5in}
\caption{\label{fig3} Distributions of $M_{CT}$ values for SUSY signal
events for ten different values of the cut on the measured $p_T$ of
the $j_1j_2+E_T^{miss}$ CoM frame (Eqn.~\ref{eqn20}), ranging from 200
GeV (top) to 20 GeV (bottom) in 20 GeV steps. In contrast to
Fig.~\ref{fig4-4} the jet multiplicity cut has been relaxed to require
at least two jets and a logarithmic $y$-axis has been used to aid
comparison of shapes of distributions.}}

In order to study the dependence of the shape and position of the
$M_{CT}$ end-point in Fig.~\ref{fig4-4} on the cut on the measured
$p_T$ of the $j_1j_2+E_T^{miss}$ CoM frame (Eqn.~\ref{eqn20}), the
$M_{CT}$ distributions of SUSY signal events passing progressively
harder $p_T$ cuts were generated. Due to the strong correlation
between the di-jet multiplicity cut and the $p_T$ cut the former cut
was relaxed to require {\it at least} two jets, with the two hardest
jets being used to calculate $M_{CT}$. The resulting distributions are
plotted in Fig.~\ref{fig3} for ten different values of the $p_T$ cut
ranging from 200 GeV (top) to 20 GeV (bottom). The effect of the cut
on the SUSY signal is to sharpen the end-point at the expense of
statistics. This sharpening of the end-point is caused both by
limitation of the transverse boost of the $\sqr\sqr$ CoM frame and by
rejection of high multiplicity non-$\sqr\sqr$ events in which two jets
from the same SUSY decay chain are selected to calculate $M_{CT}$. It
should also be noted that a harder $p_T$ cut rejects more SM background
events, especially QCD multijet events.

\section{Extension to measurement of individual particle masses}
\label{sec5}

So far we have shown that we can obtain an end-point in the
distribution of a quantity calculated from visible decay product
transverse momenta which depends on an analytical combination of
masses. In principle it is possible to use the position of this
end-point to measure the individual masses $m(\delta)$ and
$m(\alpha)$. The key to this is recognising that $\poi$ depends upon
both the unknown masses $m(\delta)$ and $m(\alpha)$ and the visible
masses $m(v_i)$. If one requires that $m(v_1)=m(v_2)=m(v)$ then it can
be shown from Eqns.~\ref{eqn5} and ~\ref{eqn14} that for given $m(v)$
the bound on $M_{CT}$ is given by:
\begin{equation}
M_{CT}^{max} = \frac{1}{m(\delta)}m^2(v) + \poo.
\label{eqn17}
\end{equation}

Consequently if events could be found in which $m(v_1)$ and $m(v_2)$
were non-zero and equal, for instance by accurately combining products
from several decays in a multi-step chain, then the position of
$M_{CT}^{max}$ would depend linearly on $m^2(v)$ with a gradient of
$1/m(\delta)$ and intercepting the ordinate at $\poo$. The gradient of
a linear fit to this bound therefore measures $m(\delta)$
independently of $m(\alpha)$, while the intercept then allows
$m(\alpha)$ to be constrained. As an aside it is interesting to note
that in this case $M_{CT}$ is also bounded from below by:
\begin{equation}
M_{CT}^{min} = 2\sqrt{m^2(v)}.
\label{eqn18}
\end{equation}
and for exclusive decay chains $m^2(v)$ is bounded from above by a
separate analytical combination of masses identical to that used by
the end-point method discussed in Section~\ref{sec1}.

The above technique should work in principle however it is likely to
be very difficult to implement in practice. The first difficulty is
connected with unambiguously associating decay products with SUSY
decay chains. One possible approach would involve focusing on specific
exclusive decay chains, using the values of invariant masses of
combinations of decay products to associate decay products to chains
\cite{Nojiri:2007pq}. Unfortunately however the low acceptance of such
exclusive selections is likely to prevent successful application of
this technique before significant quantities of data have been
acquired.

A second approach involves inclusive selection of SUSY events with
multiple visible decay products, and use of a kinematic algorithm to
approximately associate decay products to
chains. Fig.~\ref{fig7}(left) shows the result of an attempt at
applying this approach to SPS1a events {\it with ISR turned off}. Here
decay products have been associated to chains by requiring that
$\max[m^2(v_1),m^2(v_2)]$ is minimised. Events have been selected by
requiring four jets and no leptons, and the mass-squared equality
requirement mentioned above has been imposed by requiring that the
asymmetry in $m^2(v_1)$ and $m^2(v_2)$ is less than 10\%. An
additional cut requiring the rapidity difference between $v_1$ and
$v_2$ to be greater than 1.0 has also been applied to reduce
combinatorics. 

\FIGURE[ht]{
\epsfig{file=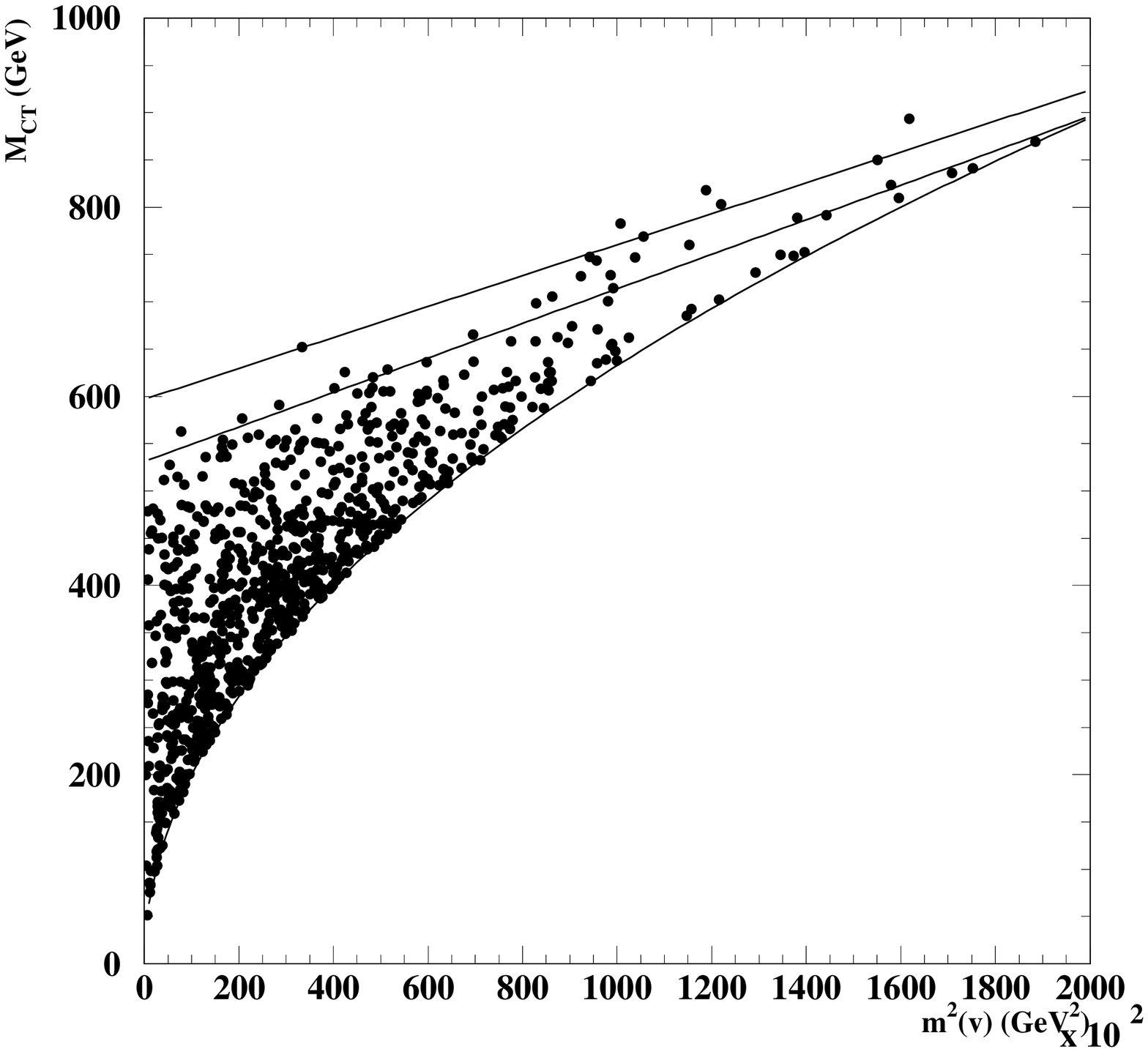,height=2.9in}
\epsfig{file=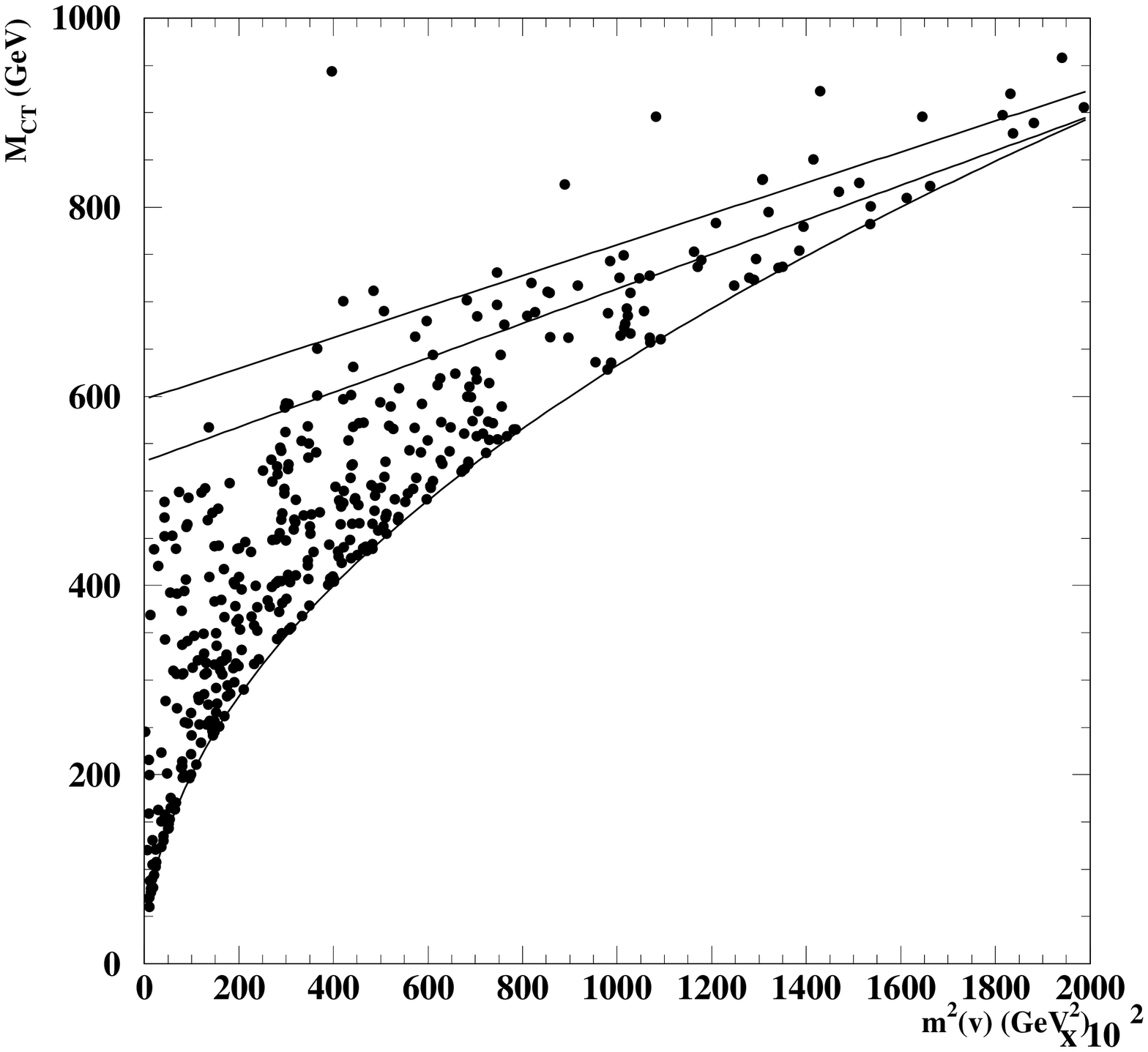,height=2.9in}
\caption{\label{fig7} Distribution in the $M_{CT}-m^2(v)$ plane of
SPS1a events with ISR turned off (left) and on (right). The top
(straight) line represents the expected dependence from
Eqn.~\ref{eqn17} of $M_{CT}^{max}$ on $m^2(v)$ for events containing
$\gl$ pair production. The middle (straight) line represents the
equivalent dependence for events containing $\tq$ pair production. The
bottom line represents the expected dependence of $M_{CT}^{min}$ on
$m^2(v)$ given by Eqn.~\ref{eqn18}.}}

Fig.~\ref{fig7}(left) shows that with these cuts events generally lie
below the expected upper bounds on $M_{CT}$ for $\gl$ decays (top
line) or $\tq$ decays (middle line), although some combinatorial
contamination is visible above the $\gl$ bound. The lower bound on
$M_{CT}$ given by Eqn.~\ref{eqn18} is prominent (bottom
line). Fig.~\ref{fig7}(right), obtained with SPS1a events {\it with
ISR turned on}, illustrates the further difficulty of using this
approach however. The inclusion of ISR jets in $v_1$ and/or $v_2$ can
artificially lower $m^2(v)$ and hence generate false configurations
which strongly violate the upper bounds on $M_{CT}$. A related effect
was noted previously in connection with the stransverse mass technique
in Ref.~\cite{Lester:2007fq}. Clearly much more work is needed before
this technique can be used practically to measure accurately
independent particle masses.

\section{Conclusions and directions for future work}
\label{sec6}
This paper has shown that by constructing a kinematic quantity
invariant under contra-linear equal magnitude boosts in the laboratory
transverse plane a simple analytical combination of the masses of
pair-produced particles and their invisible decay products can be
constrained at hadron colliders such as the LHC. It was shown that in
principle these techniques may be used to measure the masses of such
particles independently, although in practice this seems to be very
difficult.

The study described in this paper suggests several directions for
future work. These include:
\begin{itemize}
\item The experimental simulation study of Section~\ref{sec4} should
be repeated with more realistic full experiment-specific simulation of all
Standard Model and SUSY backgrounds to demonstrate conclusively the
feasibility of these techniques when applied to $\sqr\sqr$ events. The
feasibility of $\sls$ mass measurement with $\sls\sls$ events should
also be studied.
\item Further work assessing the feasibility of measuring independent
particle masses using the technique outlined in Section~\ref{sec5} is
required, focusing in particular on optimising the experimental
assignment of decay products to decay chains, and rejection of ISR
jets.
\end{itemize}

\section*{Acknowledgements}
DRT wishes to thank Giacomo Polesello for the simulated data samples
and analysis framework first used to investigate the ideas discussed
in this paper and for comments on an early draft. He also wishes to
thank Mihoko Nojiri and Elizabeth Winstanley for comments and Alan
Barr, Claire Gwenlan and Chris Lester for discussions. DRT wishes to
acknowledge the Science and Technology Facilities Council (STFC) for
support.

\end{document}